\documentclass{aa}
\newcommand{\be}{\begin{equation}}
\newcommand{\ee}{\end{equation}}
\usepackage{graphicx}

\begin{document}

\title{Two populations among the metal-poor field RR Lyrae stars}
\author{Borkova T.V., Marsakov V.A.}
  \institute{Space Research Department at Rostov State University, \\
            and Issac Newton Institute of Chili Rostov-on-Don Branch
            Stachki 194, Rostov-on-Don, 344090, Russia \\
  \email {borkova@rsusu1.rnd.runnet.ru, marsakov@ip.rsu.ru}}
\offprints{V.\ Marsakov}
\date{received \ 17 April 2002  ; accepted \  17 September 2002}

\abstract{
We compute the spatial velocity components and the 
galactic orbital elements for 209 metal-poor $([Fe/H] < -1.0)$ 
RRLyrae (ab) variable stars in the solar neighborhood using 
proper motions, radial velocities, and photometric 
distances available in the literature. The computed orbital 
elements and published heavy element abundances are 
used to study relationships between the chemical, spatial, 
and kinematical characteristics of nearby field RR Lyrae variables. 
We observe abrupt changes in the stellar spatial and 
kinematical characteristics when the peculiar velocities 
relative to the local standard of rest cross the threshold 
value, $V_{\rm pec}\approx 280$ \mbox{km\,s$^{-1}$}. This provides 
evidence that the general population of metal-poor RRLyrae stars is 
not uniform, and includes two spherical subsystems occupying 
different volumes in the Galaxy. Based on the agreement between 
typical parameters of corresponding subsystems of field RRLyrae 
stars and of the globular clusters, studied by us earlier, we 
conclude that metal-poor stars and globular clusters can be 
subdivided into two populations, but using different criteria for 
stars and clusters. We suppose that field stars with velocities 
below the threshold value and clusters with extremely blue 
horizontal branches form the spherical, slowly rotating subsystem of 
the proto-disk halo (related by its origin to the Galactic 
thick disk). It has a negligible, but non-zero, vertical 
metallicity gradient. Field stars with fast motion and 
clusters with redder horizontal branches constitute the 
spheroidal subsystem of the accreted outer halo, with is
approximately two times larger in size than the first subsystem. 
It has absolutely no metallicity gradients, most of its stars 
have eccentric orbits, many stars display retrograde motion in the 
Galaxy, and their ages are comparatively low, supporting 
the hypothesis that the objects in this subsystem have an
extragalactic origin. 

}

\maketitle


\section{Introduction}

The presence of two different populations with separate 
histories in the metal-poor halo was suggested by Hartwick 
(1987). He showed that model of the dynamics of RR Lyrae 
variables with metallicities $[Fe/H]<-1.0$ required two components: 
one spherical and a somewhat flattened component that is dominant at 
galactocentric distances less than the radius of the solar circle. 
The idea that there are two subsystems in the metal-poor halo has already
been addressed in several investigations focused on globular clusters.
It turned out, that globular clusters present a distinctive inner feature 
(the morphology of their horizontal branches), which  makes it 
possible to distinguish the individual metal-poor clusters in 
terms of different halo subsystems. It was found that halo clusters, 
that have redder horizontal branches for a given metallicity (i.e.
with horizontal branches showing a considerable number of stars on the
low-temperature side of the Schwarzschild gap), are predominantly 
outside the solar circle. Furthermore, they exhibit larger velocity 
dispersion, slower circular velocity (a significant number having 
retrograde orbits), and are, on average, younger than clusters 
with extremely blue horizontal branches, which are concentrated 
within the solar circle (Da Costa \& Armandroff
1995; Borkova \& Marsakov 2000). Note that relative ages for
globular clusters were determined by different authors with the 
help of observed high-precision colour-magnitude diagrams and 
theoretical isochrones. Therefore, these ages are independent of 
horisontal branch morphology, and the relative youth of some 
metal-poor globular clusters is beyond doubt (see compilative 
catalogue of homogeneous age dating of 63 globular clusters by 
Borkova \& Marsakov (2000)). The explanation suggested for 
the difference between these two populations was that the subsystem 
of clusters with extremely blue horizontal branches 
(i.e. the older halo) formed together with the Galaxy as a whole, 
whereas the clusters of the younger halo subsystem formed from 
fragments captured by the Galaxy from intergalactic space at later 
stages of its evolution (Zinn 1993). Recent observations (Ivezic 
et al. 2000, Vivas et.al. 2001) suggest strong evidence in
favor of the hypothesis that star and also globular clusters of 
the outer halo are the debris left over from the accretion of dwarf
galaxies. Unfortunately, it is impossible now to ascribe concrete 
nearest field RR Lyrae stars to some subsystem, because we do not 
know any intrinsic distinctive quality for them.

Some papers (Chiba \& Yoshii 1998; Martin \& Morrison 1998; 
Dambis \& Rastorguev 2001) present detailed studies of the kinematics 
of RR Lyrae stars in the solar neighborhood. The stars were 
assumed to form only one subsystem in a metal-poor halo. We 
use spatial velocities and computed elements of galactic orbits 
as our main criteria for isolating subsystems (because of the 
local, near-solar position of the studied RR Lyrae stars). Our 
wish is to investigate relationships between physical, chemical, 
spatial and kinematical characteristics of RR Lyraes stars in 
each metal-poor subsystem, determine the characteristic 
parameters of these subsystems, and compare them with the 
parameters of similar subsystems of globular clusters.


\section{The Data}
\label{sectdata}

We used for this study the largest catalogue of RR Lyrae 
variables, compiled by Dambis \& Rastorguev (2001). 
The catalogue contains 262 stars with published photoelectric 
photometry, metallicities, radial velocities, and absolute 
proper motions. Dambis \& Rastorguev (2001) 
used the  proper motions  from Hipparcos, PPM, NPMI, and 
Four-Million Star Catalog (Volchkov et al. 1992); 
the proper motions from the last three ground-based catalogs have 
been reduced to the Hipparcos system. The metallicities were used 
mainly from Layden (1994) and Layden et al. (1996), whose 
metallicity scale is in conformity with metallicity scale of 
globular clusters (Zinn \& West 1984). Dambis and Rastorguev, 
(2001) used radial velocities mainly from the paper of Fernley et al. 
(1998), and Solano et al. (1997), and main magnitude $<V>$ - 
mainly from Fernley et.al. (1998). Bearing in mind the large 
relative errors of trigonometric parallaxes for distant objects, 
we chose the photometric distance scale of Dambis \& 
Rastorguev (2001), assuming $M_{\rm v}(RR)=0.26[Fe/H]+ 1.17$. 
For each star, we computed the spatial velocity components 
in the cylindrical coordinates and the orbital elements using 
the Galaxy model from Allen \& Santillan (1991), which 
includes a spherical bulge, a disk, and an extended massive 
halo. The model assumes the galactocentric distance of the 
Sun to be $R_\odot=8.5$ kpc and the Galactic circular rotation 
velocity at the solar distance to be $V_{\rm rot}=220$
\mbox{km\,s$^{-1}$}. The orbital elements were computed by 
modeling five complete revolutions around the galactic 
center for each star. The most informative quantities are: 
$Z_{\rm max}$ - the maximum height of the star above the 
galactic plane, $R_{\rm a}$ - the orbital apogalactic radius; 
$R_{\rm p}$ - the orbital perigalactic radius, and the 
eccentricity, $e=(R_{\rm a}-R_{\rm p})/(R_{\rm a}+R_{\rm p})$. 
A choice of stars based solely on their variability type 
and their visible magnitude ensures an absence of kinematical 
selection effects in the catalogue used. Our final sample of 
the metal-poor RR Lyrae variables contains 209 stars with 
metallicity $[Fe/H]<-1.0$. This simple metallicity criterion 
eliminates objects that belong to the thick disk subsystem of 
the Galaxy (see substantiations of Borkova \& Marsakov (2002)).


\section{Criteria to select the subsystems}
\label{sectcriterion}

\begin{figure*}
  \centering
  \includegraphics[angle=0,width=16cm]{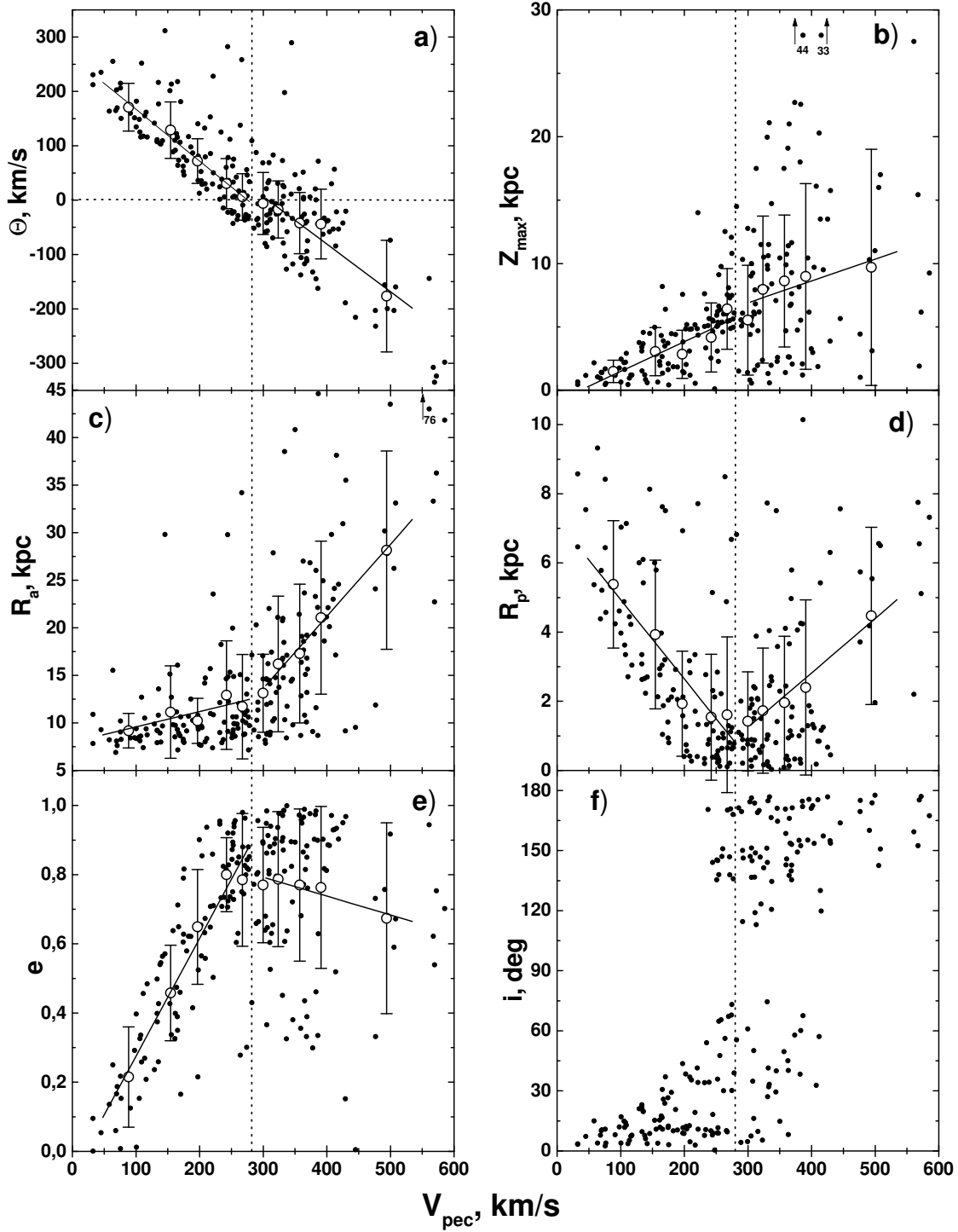}
   \caption {
The relations between the peculiar velocity 
relative to the local standard of rest, $V_{\rm pec}$, and other 
characteristics of the RR Lyrae variables. Large open circles 
with error bars are mean values and dispersions of the corresponding 
parameters in narrow intervals $V_{\rm pec}$. The straight 
lines are the least-square fits for the stars lying both to 
the left and on the right of $V_{\rm pec}\approx 280$ \mbox{km\,s$^{-1}$} 
(the vertical dotted lines). An abrupt change in the relations and 
the dispersions near this threshold level is apparent in all panels.}
       \label{fig1}
\end{figure*}

It is much more difficult to identify objects that have an 
extragalactic origin, i.e. those belonging to the 
accreted halo. According to the hypothesis that the  
protogalaxy collapsed monotonically from the halo to the disk, 
suggested by Eggen et al. (1962), stars genetically related 
to the Galaxy cannot have retrograde orbits. Only the 
oldest halo stars may be an exception, since they could have retrograde 
orbits due to the natural initial velocity dispersion of 
protostellar clouds. On the other hand, some stars formed from
extragalactic fragments and captured by the Galaxy may have a prograde
orbits. In any case, such stars should have fairly large peculiar 
spatial velocity relative to the local standard of rest, 
$V_{\rm pec}$. Fig.\ \ref{fig1}a displays the relation between 
the peculiar velocity (assuming $(U_o, V_o, W_o)$=(-10,10,6)
\mbox{km\,s$^{-1}$} and the azimuth (circular) velocity component, 
$\Theta$, for the RR Lyrae stars of our sample. The diagram 
shows that there is a transition from prograde to retrograde 
orbits around the galactic center near $V_{\rm pec}\approx 280$ 
\mbox{km\,s$^{-1}$}. We also observe an abrupt increase of the 
dispersion of the circular velocity component and a break 
of the dependence of circular velocity as a function of peculiar 
velocity at the same place (see the dispersion bars and the 
regression lines in the diagram). Fig.\ \ref{fig1}b displays a 
significant increase in the scatter of stars in $Z_{\rm max}$ 
when crossing the same threshold peculiar velocity. The abrupt 
change in the apogalactic radii of the stellar orbits is even 
more evident (Fig.\ \ref{fig1}c). First, the mean value of Ra 
remains much the same and when peculiar velocity crosses the 
threshold level, the mean value of apogalactic distances 
monotonically increase and their scatter sharply extends. We 
can see in Fig.\ \ref{fig1}d that the perigalactic 
distances also demonstrate the break of their relations in 
the vicinity of the same threshold peculiar velocity. The orbital 
eccentricities (Fig.\ \ref{fig1}e) not only abruptly 
increase their dispersion when crossing the same point, 
they also demonstrate different relations. First, the 
orbital eccentricities increase almost linearly with 
the peculiar velocities, reaching a maximum near the 
threshold velocity level. With further increase of $V_{\rm pec}$, 
the mean and the scatter of the eccentricities do not 
change, within errors. Fig.\ \ref{fig1}f shows that RR Lyrae 
stars with velocities near the threshold level might have any 
orbital inclination up to orthogonal to the galactic plane, 
whereas the range of  "permitted" inclinations 
continuously becomes narrower when velocity moves away from 
this threshold level on both sides.  For these reasons, 
we adopt $V_{\rm pec}>280$ \mbox{km\,s$^{-1}$} as the 
critical value for distinguishing stars of the outer accreted
halo. Here we suppose that the stars with lower peculiar 
velocities have a galactic origin, and belong to the proto-
disk halo subsystem. Apparently, this kinematical criterion is 
not entirely unabbiguous: some stars of the proto-disk halo may 
have larger residual velocities. Evidence for this is provided, 
in particular, by the increase in the stellar density immediately
to the right of the thereshold velocity level in our diagrams. 
However, we decided to retain a simple criterion, in order not 
to artificially confuse the situation.

Note again that the principal criterion distinguishing globular 
clusters of the accreted halo is their redder horizontal 
branches compared to clusters of the proto-disk halo. Moreover, 
empirical evidence suggests that the morphological structure of 
the horizontal branches is related to the mean period of 
RR Lyrae variables in metal-poor clusters (vanAlbada \& Baker 
1972). We investigate the detailed data for RR Lyrae in globular 
clusters from Clement et al. (2001), and show that the redder the horizontal
branch, the lower the mean period with a height correlation coefficient 
$r=0.8\pm 0.1$. It means that the mean period for RR Lyrae in 
clusters can be used to divide this cluster into 
\begin{table*}
\centering
\caption[]{\bf{Characteristics of metal-poor subsystems of field RR Lyrae
           variables and Globular Clusters}}
         \label{table}
\begin{tabular}{|l|c|c|c|c|}
\hline  \bf{Characteristics}  & \multicolumn{2}{c|}{\bf {Proto-disk halo}} & 
\multicolumn{2}{c|}{\bf {Accreted halo}} \\
\cline{2-5}
       &\bf{RR Lyr} & \bf{GC} & \bf{RR Lyr} & \bf{GC} \\
\hline 
       $<[Fe/H]>$, dex & -1.56$\pm$ 0.04& -1.71$\pm$ 0.05& 
                         -1.62$\pm$ 0.03& -1.60$\pm$ 0.07 \\
$\sigma_{[Fe/H]}$, dex &  0.33$\pm$ 0.05&  0.26$\pm$ 0.03& 
                          0.32$\pm$ 0.02&  0.35$\pm$ 0.05 \\
      $<\Theta>$, km/s &    93$\pm$ 8   &    77$\pm$ 33 & 
                           -51$\pm$ 11  &   -23$\pm$ 54 \\
    $\sigma_{\Theta }$,
                  km/s &    83$\pm$ 6   &   129$\pm$ 19 &
                           106$\pm$ 8   &   140$\pm$ 18 \\
                 $<e>$ &  0.57$\pm$0.03 &  0.53$\pm$0.06 &
                          0.76$\pm$0.02 &  0.59$\pm$0.06 \\
   lim$R_{\rm a}$, kpc &       18       &        10      & 
                               42       &        20     \\
lim $Z_{\rm max}$, kpc &       9        &        10      & 
                               22       &        20     \\
            $Z_0$, kpc &   2.6$\pm$0.2 *  &   2.5$\pm$0.5  & 
                           6.1$\pm$0.3  &   8.5$\pm$1.5  \\
$\rm grad_{\rm R}[Fe/H], 
          \rm kpc^{-1}$& -0.00$\pm$0.01 & -0.03$\pm$0.02 & 
                         -0.00$\pm$0.01 & -0.03$\pm$0.01 \\
$\rm grad_{\rm Z}[Fe/H],
          \rm kpc^{-1}$& -0.02$\pm$0.01 & -0.03$\pm$0.02 & 
                         -0.00$\pm$0.01 & -0.03$\pm$0.01  \\
\hline
\end{tabular}
\begin{list}{}{}
  \item[Note.*] - The gradient of the proto-disk halo is strongly 
                distorted by the selection which depends on only
                nearest stars.
\end{list}
\end{table*}
subsystems. Therefore, it is interesting to test whether 
the variability period of single field RR Lyrae star may be 
adopted as well. To test this for field stars, we 
computed the mean periods of stars in a narrow metallicity range
$(-1.7<[\mathrm{Fe/H}]<-1.2$) for both halo subsystems. Each 
sample contained about sixty stars. It turns out that their mean 
periods are the same. Thus, the variability period of the field 
RR Lyrae stars cannot serve as an additional (internal) criterion 
to divide them into different subsystems of the metal-poor halo. 

\section{Properties of stars in the subsystems}
\label{sectproperties}

\begin{figure*}
  \centering
  \includegraphics[angle=0,width=16cm]{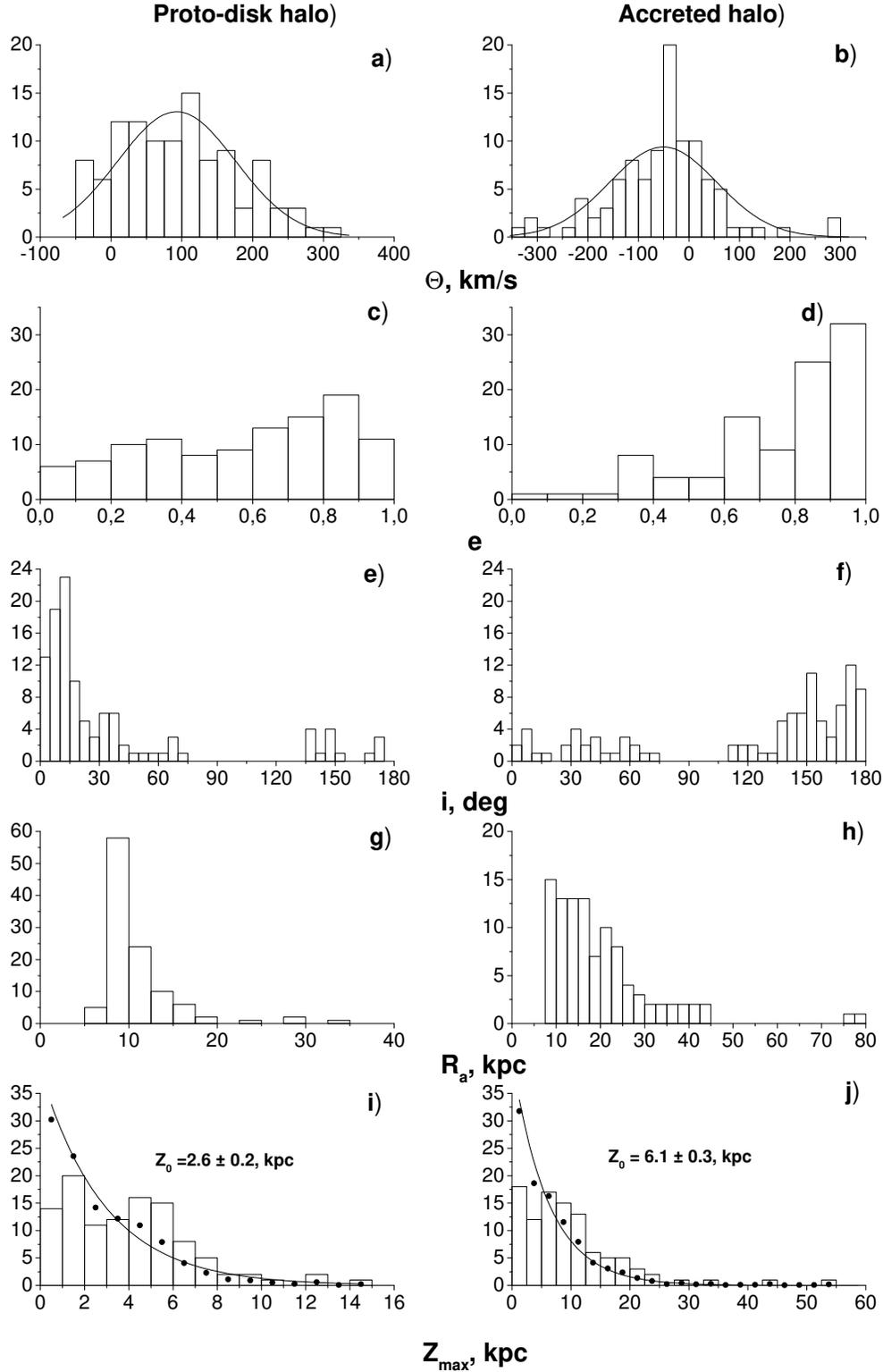}
   \caption {
Distributions of orbital elements for the RR Lyrae stars. Result 
for stars with $V_{\rm pec}<280$ \mbox{km\,s$^{-1}$} (left column),
$V_{\rm pec}>280$  (right colimn). The curves 
in the top row represent an approximation of the distributions of 
Gaussian functions. In bottom row, the dots are the reconstructed
distributions of RR Lyrae stars in Z (see the text for details),
and the curves approximate the reconstructed distributions with 
an exponential law (the corresponding scale heights and their 
uncertainties are indicated).
}
       \label{fig2}
\end{figure*}

Let us now compare the properties of the resulting subsystems. 
The spatial velocities of the stars can be used to obtain an estimate 
for a number of characteristics of subsystems, if we can first 
reconstruct the stars' galactic orbits. Fig.\ \ref{fig2} shows the 
distributions of RR Lyrae in the two metal-poor subsystems as a 
function of their orbital elements. The top panels present the
histograms of the rotation velocities for the sample of stars in
the proto-disk halo and in the accreted halo. It is worth nothing
that the two distributions can be properly fitted by a Gaussian 
(see the solid lines). Their maxima are separated by almost the 
dispersion (Table \ref{table}). The second row of graphs in 
Fig.\ \ref{fig2} shows the corresponding distributions in orbital 
eccentricities. Here the characters of the histograms are obviously 
very different. All eccentricities are present in approximately 
equal number in the proto-disk halo (only a small excess is 
seen towards the high eccentricity side). Stars with very eccentric orbits 
prevail in the outer halo (where almost two thirds of all 
stars have $(e>0.8)$). The next row (Fig.\ \ref{fig2}e,f) 
presents distributions of orbital inclinations. The 
stars of both subsystems can have any orbital 
inclination. In both cases, the number of stars strongly 
increases with decreasing inclination; however, this is true 
for stars with prograde orbits in the proto-disk halo, and 
for those with retrograde orbits in the outer halo. We must
bear in mind that the deficiency of stars with large orbital 
inclinations is largely due to the kinematical selection 
effect imposed on the sample of nearby stars. The vertical
components of the spatial velocities of such stars in the 
solar neighborhood should be comparable to the galactic 
rotational velocity at this distance. Thus, the probability
of their presence here is very low. 

\begin{figure*}
  \centering
  \includegraphics[angle=0,width=18cm]{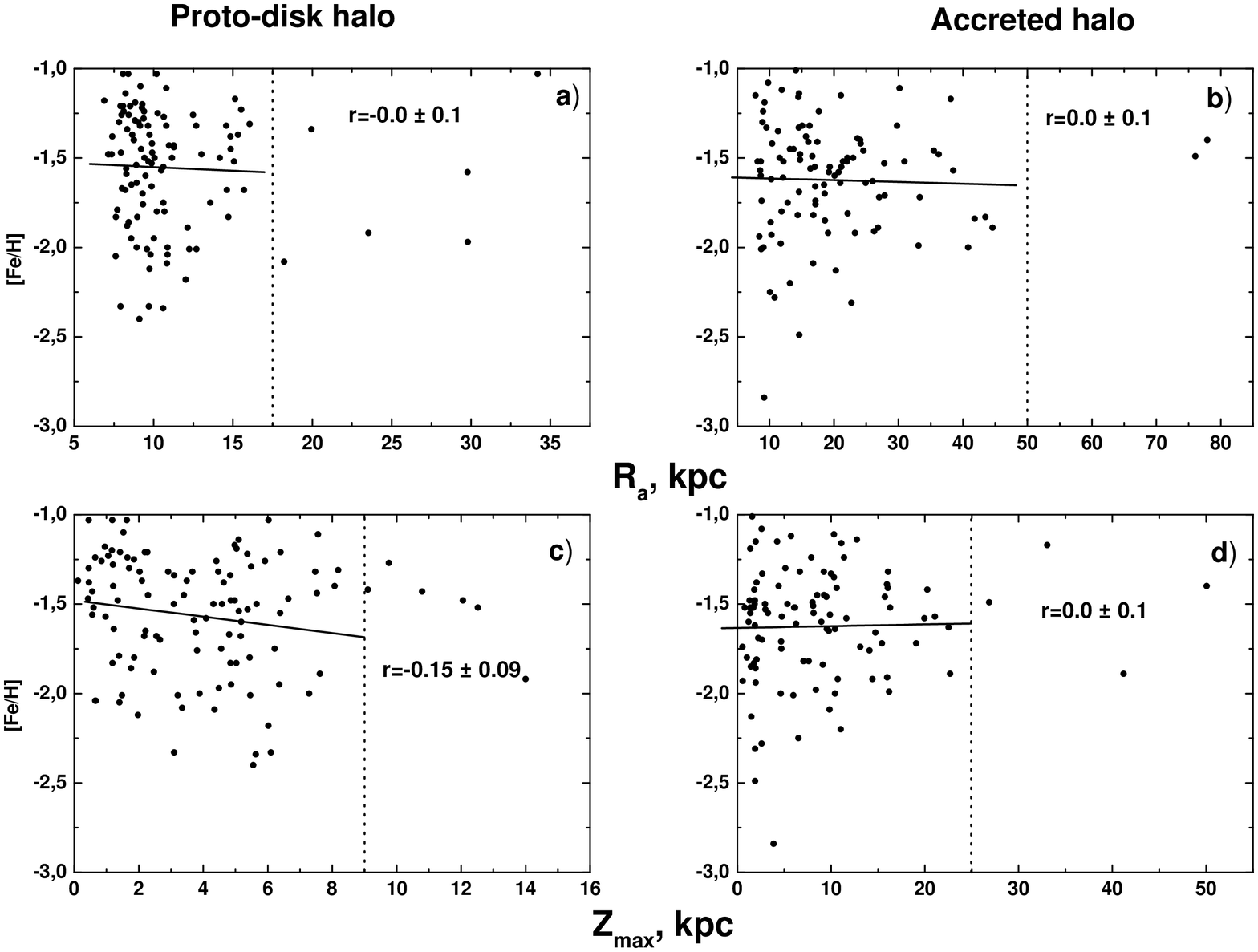}
   \caption {
Metallicities of the RR Lyrae stars versus apogalactic 
distance (a,b,) and maximum height above the galactic plane (c,d). 
The straight lines are least-square fits. The slopes of the lines 
determine the metallicity gradients. The most distant stars in each 
panel (on the right of the dotted lines) were rejected from our 
computations. The correlation coefficient and their uncertainties 
are given.}
       \label{fig3}
\end{figure*}

The fourth row of the histograms (Fig.\ \ref{fig2}g,h) can be 
used to estimate radial sizes of the subsystems and the fifth
row (Fig.\ \ref{fig2}i,j), their vertical sizes. To quantitatively 
estimate the outer sizes of the subsystems based on these 
distributions, we reject the five most distant points in each 
histogram. Therefore, we simultaneously remove the largest 
uncertainties in the determination of the orbital elements and
avoid possible errors in assigning some stars in our sample to a 
particular subsystem. Such estimates indicate that the size of
the outer halo is approximately a factor of two larger than that
one of the proto-disk halo (Table\ \ref{table}). The sizes of 
the subsystems in the direction perpendicular to the galactic 
plane also differ drastically (Table\ \ref{table}).

It is obviously not correct to compute the scale height using  
$Z_{\rm max}$, since all stars of a subsystem cannot 
simultaneously be located at the the highest points of their 
orbits. To reconstruct the "real", instantaneous Z-distribution 
for all stars, we must "spread" each star over its orbit from 
$-Z_{\rm max}$ to $+Z_{\rm max}$ in proportional to the 
probability density of its location at different Z. (This 
operation also smoothes fluctuations in the histogram due to 
the limited number of stars in the sample.) This probability 
density can easily be found from the computed orbit of the stars. 
The details of the procedure were written by Marsakov \& 
Shevelev (1995). The filled dots in the histograms of the 
bottom row (Fig.\ \ref{fig2}i,j) are the final reconstructed 
distributions in Z. In other words, this is how the stars will 
be distributed in height after some time, if they are rundomly 
distributed in their orbits. The solid curves in Fig.\ \ref{fig3}i,j are 
least-square approximations of the reconstructed distributions 
using an exponential law, 

  $${\large n=\alpha \cdot e^{-Z/Z_0}},$$
   
where $Z_0$ is the scale height (the corresponding $Z_0$ values 
are indicated in Fig.\ \ref{fig3}i,j and in Table\ \ref{table}).

Let us now consider the metallicity gradients in the subsystems. 
Fig.\ \ref{fig3} displays the $R_a - [\mathrm{Fe/H}$] and 
$Z_{\rm max} - [\mathrm{Fe/H}]$ diagrams. The straight 
lines are the least-square regressions. Assuming that the stars 
are born near the apogalactic radii of their orbits, the slopes 
of these lines reflect the initial radial and vertical 
metallicity gradients for the subsystems. To increase the accuracy
of the gradient estimates, we rejected the most distant data points 
in each case (see the vertical dotted lines in the diagrams). 
The resulting gradients are presented in Table 1 and the 
correlation coefficients in the corresponding diagrams. 
Only the vertical gradient in the proto-disk halo exceeds the
error, but the correlation coefficient given in Fig.\ \ref{fig3}c 
tends to zero if we add the rejected distant stars. Note that
RR Lyrae metallicities from Zinn \& West (1984) are estimated using
the $\Delta S$ method, which actually measures calcium abundance.
According to current evidence there is a systematic uncertainty
between the metallicity scale suggested by Zinn \& West (1984) 
and other ones (see Rutledge et al. 1997, Rey et al. 2000). 
Therefore, to check our result we have to use more reliable
metallicity determinations for field RR Lyrae stars, when sufficient
data will be available. Thus the existence of 
vertical gradient in the proto-disk halo is still an open 
question. In contrast, according to our investigation, the complete 
absence of a radial gradient in the proto-disk halo and of both 
gradients in the accreted halo for the RR Lyrae stars is beyond 
doubt. 

Suntzeff et al. (1991) investigate radial and vertical
metallicity gradients for the metal-poor field RR Lyrae population 
as a whole. They used the present positions of stars in selected
regions of the Galaxy in the galactocentric distance range 
4-30 kpc. The authors indicate that outside the solar circle
the metallicity gradients are zero as a function of $R_{\rm a}$
or $|Z|$. This result is in agreement with gradients in
our accreted halo estimated on $R_{\rm a}$ and $Z_{\rm max}$.
Inside the solar circle they find the radial gradient of 
$-0.06 \pm 0.02$ dex \mbox kpc$^{-1}$, which is steeper than 
that of our proto-disk halo for field RR Lyraes (but is in 
agreement for globular clusters). This discrepancy is due 
to the lack of observed RR Lyraes inside the solar circle. The 
vertical gradient in [Fe/H] they  estimate as roughly -0.05 
dex \mbox kpc$^{-1}$ out to 3 kpc. Our value for the proto-disk 
halo is larger because Suntzeff et al. (1991) did not divide 
the metal-poor stars into subsystems. So we may conclude that
computed orbital elements for the nearest stars are good enough 
for estimation of metallicity gradients in the much larger
distance range (but not inside the solar circle.)


\section{Discussion and conclusion}
\label{sectdiscussion}

Here, we will compare the characteristics of the metal-poor RR Lyrae 
subsystems derived in this study to the parameters of 
corresponding subsystems of globular clusters from Borkova \& 
Marsakov (2000), since only globular clusters were distinguished
according to an intrinsic, physical parameter rather than
interrelated spatial and kinematical criteria. The parameters of 
the metallicity distributions of the corresponding halo subsystems 
differ somewhat. In particular, the mean metallicity of the 
proto-disk halo derived from the globular clusters is lower than 
that of the outer halo. The metallicity dispersion is also lower. 
The field stars show the opposite pattern (Table\ \ref{table}). 
In all cases, however, the differences are comparable to the formally 
computed uncertainties, indicating that any conclusions about 
differences between these parameters have low statistical 
significance. The vertical gradients in the proto-disk halo 
for both the RR Lyrae stars and the globular clusters 
almost coincide, but the radial gradients differ (see 
Table\ \ref{table}). In the accreted halo, both gradients 
are absent for the field RR Lyrae stars but are non-zero for the 
clusters. However, both gradients for the globular clusters in 
the accreted halo are due exclusively to metal-richer objects 
close to the galactic center ($R\sim 7$ kpc). Distant RR Lyrae
were not included in our sample. In any case, values of all 
corresponding gradients coincide within the uncertainties. 
In this study, we have identified the halo subsystems based on
spatial velocity. Therefore, the differences between the proto-disk 
halo and the accreted halo in any kinematical parameter for 
field stars should be more prominent. Indeed, while the 
difference between the orbital velocities for the globular
clusters of the proto-disk halo and of the accreted halo is  
$\sim 100$ \mbox{km\,s$^{-1}$} this difference is approximately 
40\% higher for the RR Lyrae variables (see Table\ \ref{table}). 
The velocity dispersions for the subsystems of globular 
clusters are obviously overestimated due to the large 
distance uncertainties and, as a result, are much higher 
than the values for field RR Lyrae stars. The mean 
eccentricities in the proto-disk halo subsystems are the same, 
whereas in the outer halo the eccentricities are, on average, 
higher for the field stars, as expected (note that the 
proper motions and, hence, orbital eccentricities are known 
only for a small number of clusters, and with large  
uncertainties). The radial size of the proto-disk halo is  
approximately a factor of 1.8 larger for the field stars than 
for the clusters, whereas the two scale heights were the same 
within the errors. Recall that we can estimate the radial 
sizes of RR Lyrae subsystems only from their maximum 
distances from the galactic center, which leads to appreciable 
overestimates of these sizes. The radial and vertical sizes of the 
outer accreted halo subsystem of field stars are naturally 
the largest, and are in reasonable agreement with the 
corresponding sizes for the subsystem of globular clusters. 
Note that, in order to obtain correct estimates of sizes 
of galactic subsystems based on data for nearby stars, it is 
necessary to take into account the kinematical selection 
effect, which leads to a deficiency of stars with large 
$R_{\rm a}$ and $Z_{\rm max}$ in the solar neighborhood.

Thus, we found that the generally good agreement between the 
characteristics of corresponding subsystems of field RR 
Lyrae stars and globular clusters, distinguished using different 
criteria, shows that both populations are not uniform. Both the 
clusters and field stars belong to two spherical subsystems 
of the Galaxy: the inner proto-disk halo (related to the disk 
by its origin) and the outer accreted halo. The collected 
results indicate that this subsystem is characterized by  
large size, an absence of appreciable metallicity gradients, 
predominantly large orbital eccentricities, a large number of 
objects on retrograde orbits, and, on average, younger ages 
for its objects, supporting the hypothesis that objects in 
this subsystem have an extragalactic origin.


\begin{acknowledgements}

This study was supported by the RFBR (projects 00-02-17689 and 
02-02-06911).
\end{acknowledgements}

\end{document}